\documentclass[3p, twocolumn]{elsarticle}

\usepackage{amsmath}
\usepackage{amssymb}
\usepackage{hyperref}

\begin{document}

\title{No probability loophole in the CHSH}
\author[leiden]{Richard D.~Gill}
\ead{gill@math.leidenuniv.nl}
\address[leiden]{Mathematical Institute, University of Leiden, Netherlands. \url{http://www.math.leidenuniv.nl/~gill}}

\begin{abstract}
\hyperref[han]{Geurdes (2014, \emph{Results in Physics})} outlines a 
probabilistic construction of a counterexample to Bell's theorem. He gives a procedure to repeatedly sample from a specially constructed ``pool'' of local hidden variable models (depending on a table of numerically calculated parameters) and select from the results one LHV model, determining a random value $\mathcal S$ of the usual CHSH combination $S$ of four (theoretical) correlation values. He claims $\mathrm{Prob}(|\mathcal S| > 2) > 0$. We expose a fatal flaw in the analysis: the procedure generates a \emph{non-local} hidden variable model. 

To disprove this claim, Geurdes should program his procedure and generate random LHV's till he finds one violating the CHSH inequality.
\end{abstract}


\maketitle

\noindent \hyperref[han]{Geurdes (2014)} discusses the CHSH inequality within a conventional framework. I will first set down the basic definitions of that paper.

A local hidden variable model (LHV) for a model of a Bell-CHSH experiment is taken to consist of a 4-tuple $\mathcal L = (A, B, \mathcal Q, \rho)$ where $A$ and $B$ stand for the two local \emph{measurement functions}, $\mathcal Q$ stands for a \emph{quartet} or set of four pairs of experimental \emph{measurement settings}, and $\rho$ is the p.d.f.\ (\emph{probability density function}) of an underlying hidden variable. When we specify an LHV $\mathcal L$, we implicitly also fix, in a compatible way, domains of the three functions $A$, $B$, $\rho$, and a larger set of which $\mathcal Q$ is a subset: the model involves not just the pairs of settings used in the experiment but also other possible settings (Geurdes' construction requires existence of one or more settings in the \emph{model} which are not used in the \emph{experiment}). 

We must distinguish between settings in the model, and settings used in the experiment. I will denote by $\overline {\mathcal A}$ and  $\overline {\mathcal B}$ the ``large'' sets of all possible measurement settings for the two parties Alice and Bob. The pairs of settings used in the experiment will be denoted, as Geurdes does, by $\mathcal A = \{1_A, 2_A\} \subseteq \overline {\mathcal A}$ and $\mathcal B = \{1_B, 2_B\} \subseteq \overline {\mathcal B}$. From this we get $\mathcal Q = \mathcal A \times \mathcal B$ with a cardinality of four; hence the name ``quartet''. 

Geurdes denotes by $\Lambda$ the set of all hidden variable values, and makes the conventional assumptions and definition
$$A~:~\Lambda\times\overline{\mathcal A}~\to~\{-1, +1\},$$ 
$$B~:~\Lambda\times\overline{\mathcal B}~\to~\{-1, +1\},$$
$$\rho~:~\Lambda~\to~[0,\infty),$$ 
$$\int_{\lambda \in \Lambda} \rho(\lambda)\mathrm d \lambda ~=~ 1,$$
$$E(a, b) ~ = ~ \int_{\lambda \in \Lambda} A(\lambda, a) B(\lambda, b) \rho(\lambda)\mathrm d \lambda.$$
I change his notation in preferring to write $\lambda$ as an \emph{argument} rather than as a \emph{subscript}\footnote{Geurdes writes in a common informal style where $\int_{\lambda \in \Lambda} \dots \rho(\lambda)\mathrm d \lambda$ means integration over any space with respect to a non-negative measure of total mass one. What is important to note is that his proof uses only conventional manipulations with integrals. Indeed, part of his proof is computational (or numerical), all spaces constructed in his paper are Euclidean, all probability measures are either continuous (with density) or discrete, all expectation values are ordinary integrals or ordinary sums.}.

$E(a, b)$ is the theoretical correlation between Alice and Bob's outcomes according to the model $\mathcal L$, when they use settings $a$ and $b$, and we define 
$$S~=~ E(1_A, 1_B) - E(1_A, 2_B) - E(2_A, 1_B) - E(2_A, 2_B).$$
I will write $S = S_{\mathcal L}$ when it is important to emphasize its dependence on the LHV model $\mathcal L$, and similarly $E_{\mathcal L}(q)$, for $q = (x,y) \in\mathcal Q$, in order to make explicit the dependence of $E(x, y)$ on $\mathcal L$. 

Combining the four integrals on the right hand side of the definition of $S$ to one, we can express $S$ as the expectation value, when $\lambda$ is sampled from the p.d.f.\ $\rho$, of 
$A(\lambda, 1_A)(B(\lambda, 1_B)-B(\lambda, 2_B)) - A(\lambda, 2_A)( (B(\lambda, 1_B) + B(\lambda, 2_B))$.
Write the integrand compactly as $A_1(B_1 - B_2) - A_2(B_1 + B_2))$. For given $\lambda$, either $B_1 = B_2$ or $B_1 = -B_2$. Therefore either $A_1(B_1 - B_2) = 0$ or $A_2(B_1 + B_2) = 0$. In both cases, the value of the integrand reduces to $\pm 2$.  Integrating with respect to a probability measure generates a value of $S$ which of necessity lies in the interval $[-2, +2]$.

We consider $S$ as a function of $\mathcal L$. Geurdes describes a scheme whereby a LHV $\mathcal L$ is found, if one is lucky, in the combined results of a large number $N$ of independent random drawings from a pool of LHV models. The CHSH value $\mathcal S = S(\mathcal L)$ generated by this procedure is random. We have seen that for all $\mathcal L$, $|S(\mathcal L)|\le 2$. Hence $\mathrm{Prob}(|\mathcal S|\le 2) = 1$. This contradicts the main result of Geurdes (2014).

It follows that there must be a mistake in Geurdes' proof, though it could better be characterized as a ``proof outline''. The proof is difficult to follow: the attempt to reduce a complex construction to the limits of a two page ``micro-article'' has resulted in missing definitions and proof steps, ambiguities and anomalies\footnote{There are dubious claims, for instance, ``$\Pr(E_{\textrm T}\approx c) >  0 $ implies that $\Pr(E = c) >  0$''. Strangely, $\lambda=(\lambda_1,\lambda_2)$ and $\rho$ appears to be a product measure. $A$ appears to depend only on $\lambda_1$ and $B$ on $\lambda_2$ which would make $E(a, b)$ a product $E(a)E(b)$.}. The proof depends on unpublished numerical computations and does not offer rigorous error bounds.  
Fortunately, it is easy to indicate a fundamental and fatal conceptual error \dots\ \emph{in the very last lines of the proof}. 

The first lines of Geurdes' final section ``Conclusion'' summarizes the results obtained so far. The context is a CHSH-type experiment, using a sequence of setting pairs $q_n  = (x, y)_n \in\mathcal Q$ in $N$ trials numbered $n= 1, ..., N$.  At each trial, the usual Alice and Bob, assisted by a third person Carrol, perform (locally) various auxiliary and independent randomisations, leading to a realisation of a random LHV model $\mathcal L_n$ with the same quartet $\mathcal Q$. At the $n$th trial we can therefore compute a ``random'' correlation $E_{\mathcal L_n}(q_n)$. This is what Geurdes denotes $E_{\textrm{T(C)}}(x, y)_n$: the subscript T, or alternatively C, denoting two different ways, given in Geurdes' formulas (2) and (3), to compute the \emph{same} quantity $E_{\mathcal L_n}(x_n, y_n)$ (the two expressions are available due to the particular special structure of the pool of LHV models).  

Geurdes claims in the first line of his last section that he has so far proven the existence of $n_k$ such that 
$$\Pr(E_{\textrm{T(C)}}(x, y)_{n_k} = E_{\textrm{QM}}(x, y)_{n_k})~>~0$$ where $(x, y)_{n_k}$, for $k = 1, 2, 3, 4$ runs through each of the four setting pairs $q$ in $\mathcal Q$. We pick out four particular trials, each with one of the possible setting pairs. By $E_{\textrm{QM}}(x, y)$, Geurdes denotes the usual singlet correlations in the standard CHSH experiment: the first is equal to $+1/\sqrt 2$, the other three are all equal to  $-1/\sqrt 2$, leading to the standard (maximal) quantum prediction for $S$ equal to $2\sqrt 2$. (Of course it is not difficult to construct four different LHV models, say $\mathcal L_q$, one for each $q \in \mathcal Q$, such that $E_{\mathcal L_q}(q) = E_{\textrm{QM}}(q)$: we do not need a sophisticated construction to do this.)

The proof proceeds with the implicit claim that four LHV models together, each one reproducing just one of the four quantum correlations, are actually one LHV model doing the job for all four correlations simultaneously: this is what is expressed in Geurdes' equation (9). The proof of (9) consists of the bare word ``hence''. But (9) does not follow from what has preceded it at all. Geurdes does not verify that the four models chosen at trials $n_1$, $n_2$, $n_3$ and $n_4$ are the same (they cannot be: CHSH!). Geurdes thinks of his whole procedure as ``local'': the four out of $N$ selected LHV's have been generated by local procedures, for each $n$ separately. But ``so what?'' The combination of four LHV's generates a \emph{non-local} hidden variable model: after both Alice and Bob's settings $x$ and $y$ have been chosen, one selects the appropriate LHV $\mathcal L_q$, depending on the pair of settings $q = (x, y) \in\mathcal Q$, to model the outcomes of Alice and Bob's measurements with those settings only.

\section*{References}
\raggedright
\frenchspacing
\def\newref{\hangindent=0.7cm}
\parskip = 0.5cm

\newref
Han Geurdes (2024), \label{han}
A probability loophole in the CHSH. \emph{Results in Physics} \textbf{4} 81--82.
\href{http://dx.doi.org/10.1016/j.rinp.2014.06.002}{doi:10.1016/j.rinp.2014.06.002}

\end{document}